\documentclass[fleqn,12pt,twoside]{article}
\usepackage{espcrc1}

\usepackage{graphicx}
\usepackage[figuresright]{rotating}

\newcommand{\AmS}{{\protect\the\textfont2
  A\kern-.1667em\lower.5ex\hbox{M}\kern-.125emS}}

\title{ Bound states of $\Theta^+$ in nuclei}

\author{E. Oset
\address{Departamento de F\'{\i}sica Te\'orica and IFIC
 Centro Mixto Universidad de Valencia-CSIC\\
 Institutos de Investigaci\'on de Paterna, Apdo. correos 22085,
 46071, Valencia, Spain}, 
 D. Cabrera\addressmark, Q. B. Li \addressmark, V.K.~Magas\addressmark, 
 M. J. Vicente Vacas\addressmark
 }
       
\begin{document}

\maketitle

\begin{abstract}

We study the binding energy and the width of the $\Theta^+$ in nuclei,
associated to the  $K N$ and $ K \pi N$ components. The first one leads to
negligible contributions while the second one leads to a sizeable attraction,
enough to bind the $\Theta^+$ in nuclei. Pauli blocking and binding effects 
on the $K N$ decay reduce considerably the $\Theta^+$ decay width in nuclei and
medium effects associated to the $ K \pi N$ component also lead to a very small
width, as a consequence of which one finds separation between the bound levels
considerably larger than the width of the states. 

\vspace{1pc}
\end{abstract}
 
 The $\Theta^+$ exotic resonance \cite{nakano,hyodo} decays into $K N$ and the width
 could be very narrow according to studies of $K N$ and $K d $ interaction 
 \cite{Gibbs:2004ji,Sibirtsev:2004bg} .
 However, inside the nucleus this width is considerably reduced as a
 consequence of Pauli blocking. An intuitive way to see this is to realize that
 a   $\Theta^+$ at rest gives rise to $K N $ with a nucleon momentum of about
 270 MeV/c, barely above the Fermi momentum at normal nuclear matter density.
 According to this, the decay would be allowed. However, as soon as the
 $\Theta^+$ has some momentum one realizes, by boosting the CM variables to the 
 frame of the
 moving $\Theta^+$, that about half of the time one has components
 of the $N$ momentum smaller than the Fermi momentum and the decay width is reduced
 to half.  This is what one sees in the quantitative study of
 \cite{cabrera,manolo1,manolo2},
 with an extra reduction with increasing binding of the $\Theta^+$. 
 One also finds there that the real part associated to this decay in the
 nucleus is of the order of 1 MeV, in agreement with \cite{kim}, and hence
 too small to produce bound states in nuclei.
 
    The real novelty comes from the two meson cloud component. A detailed study
of the contribution of the two meson cloud component to the binding energy of
the antidecuplet to which the $\Theta^+$ is assumed to belong \cite{diakonov} 
has been done in \cite{twomeson}.  

 The assumptions done in \cite{cabrera} are:
 
1) ~The $\Theta^+$  is  $J^P=\frac{1}{2}^+$ and is a member of an antidecuplet
 to which  the $N^*(1710)$ belongs.
 
2) ~Two SU(3) invariant Lagrangians involving the smallest number of
 derivatives are introduced. In one of them the two mesons are in a vector
 state and in the other one in a scalar state.   The chosen Lagrangians  are \\

  $ {\cal L}_1 = i g_{\bar{10}} \epsilon^{ilm} \bar{T}_{ijk} \gamma^{\mu} B^j_l
(V_{\mu})^k_m \, ,$\\

   ${\cal L}_2 = \frac{1}{2 f} \tilde{g}_{\bar{10}}  \epsilon^{ilm} \bar{T}_{ijk}
(\phi \cdot \phi)^j_l B^k_m \, ,$\\

\noindent with  $V_{\mu}$ the two-meson vector current, \\

$V_{\mu} = \frac{1}{4 f^2} (\phi \partial_{\mu} \phi - \partial_{\mu}
 \phi \phi)$, \\
 
\noindent and  $T_{ijl}$, $B^j_l$, $\phi ^k_m$
$SU(3)$ tensors for the antidecuplet states, the octet of  
$\frac{1}{2}^+$ baryons and the octet of $0^-$ mesons, respectively.

The couplings are then fitted to
 the partial decay widths of the  $N^*(1710)$ into $N \rho $ and $N \pi \pi
 (s-wave, I=0) $ respectively. The resulting coupling 
constants are $g_{\bar{10}}=0.72$
and  $\tilde{g}_{\bar{10}}=1.9$. The uncertainties for these constants are
quite large with the current experimental information.

3) ~With these Lagrangians the selfenergy of the members of the antidecuplet is
 evaluated and a regularizing cut off of natural order is chosen (around 800
 MeV), which provides bindings of about 100-200 MeV to the members of the
 antidecuplet from the two meson cloud, plus a splitting of the order of 20 MeV
 among the different strangeness partners of the antidecuplet. This
splitting is about as large as the one obtained from quark correlations in most
quark models, hence its importance in a detailed study of the $\Theta^+$.
 
4) ~The same selfenergy is now evaluated in nuclear matter by dressing up
 the pion, allowing it to excite $ph$ and $\Delta h$, and also the $K$, which is 
 dressed with the selfenergy provided by Chiral Perturbation Theory. 
 Diagrammatically it can be seen in Fig.~1.
 
\begin{figure}[htb]
\begin{center}
\includegraphics[width=0.5\textwidth]{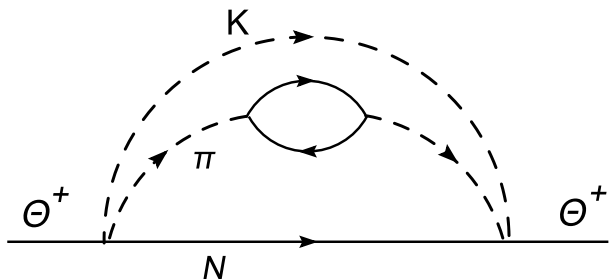} 
\end{center}
\label{med}
\caption{Selfenergy diagram due to the two meson cloud.}
\end{figure}
The results for the real and the imaginary parts can be seen in Fig. 2\\

\begin{figure}[htb]
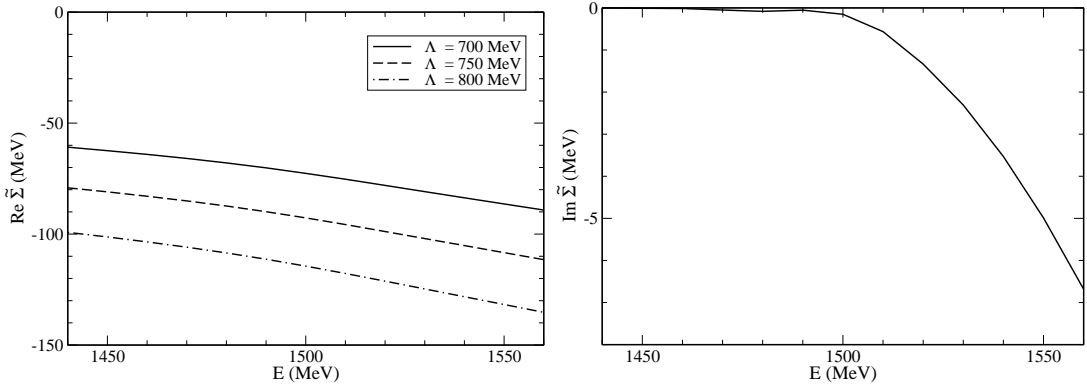

\includegraphics[width=0.45\textwidth]{ReSigma2mesondecay.eps}
\includegraphics[width=0.44\textwidth]{ImSigma2mesondecay.eps}
\caption{$\Theta^+$ selfenergy in the medium for $\rho=\rho_0$.}
\end{figure}

  What we can see in the figure is that for reasonable cut offs of the order of 
700 to 800 MeV one gets an attraction in the medium which is about 60 to 120 MeV
at normal nuclear matter density. Similarly large attractive potentials are
found  within several quark models in \cite{gerry,toki}. We can also see
in Fig. 2 the  imaginary part of the $\Theta^+$ selfenergy, 
which is cut off
independent, and is very small.  For binding energies of the order of 20 MeV it is of
the order of 3 MeV.  This width, added to the one coming from the reduced $K N$
decay in the medium, is sufficiently small to have peaks of the $\Theta^+$ bound
states perfectely identifiable.
  This is seen in Table 1. 
  
\begin{table}
	\centering
		\begin{tabular}{|c|c||c|c|}
\hline 
\multicolumn{2}{|c||}{$V=-60$ MeV $\rho/\rho_0$}&
\multicolumn{2}{|c|}{$V=-120$ MeV $\rho/\rho_0$}\\
\hline
$E_{i}$ (MeV), $^{12}C$ &
$E_{i}$ (MeV), $^{40}Ca$ &
$E_{i}$ (MeV), $^{12}C$&
$E_{i}$ (MeV), $^{40}Ca$ \\
\hline
\hline 
-34.0 (1s)&
-42.6 (1s)&
-87.3 (1s)&
-98.2 (1s)\\
-14.6 (1p)&
-30.9 (1p)&
-59.5 (1p)&
-83.3 (1p)\\
-0.3 (2s) &
-18.7 (1d)&
-32.0 (2s)&
-67.5 (1d)\\
           &
-17.9 (2s) &
-31.9 (1d) &
-65.9 (2s)\\
&
-6.3 (1f)&
-8.6 (2p)&
-50.8 (1f)\\
&
-5.6 (2p)&
-5.6 (1f)&
-48.5 (2p)\\
&
&
&
-33.5 (1g)\\
&
&
&
...\\
\hline
			
		\end{tabular}
\caption{Binding energies of the $\Theta^+$ states for two different $\Theta^+$
potentials.}		
\end{table}

As one can see, for a reasonable potential of $60 \rho/\rho_0$ $MeV$, similar to
the one of the nucleons, both in nuclei like $^{12}C$ or $^{40}Ca$, the states
are separated by an energy fairly larger than the width that we have calculated
before.  This would make a case for clear experimental observation. Suggestions
for experiments have been done using the $(K,\pi)$ reaction in nuclei
\cite{nagahiro} and there are 
plans to make the experiment at KEK \cite{imai}.\\

{\bf Acknowledgments \ }
This work is partly supported by DGICYT contract number BFM2003-00856,
and the E.U. EURIDICE network contract no. HPRN-CT-2002-00311. 
This research is part of the EU integrated infrastructure initiative
      Hadron Physics project under contract number  RII3-CT-2004-506078.

\end{document}